\documentclass[lettersize,journal,twoside]{IEEEtran}
\PassOptionsToPackage{hyphens}{url}
\usepackage{amsmath,amsfonts}
\usepackage{algorithmic}
\usepackage{algorithm}
\usepackage{array}
\usepackage[caption=false,font=normalsize,labelfont=sf,textfont=sf]{subfig}
\usepackage{textcomp}
\usepackage{stfloats}
\usepackage{url}
\usepackage{verbatim}
\usepackage{graphicx}
\usepackage{cite}
\usepackage{xcolor}
\usepackage{amsmath,amssymb,amsfonts}
\usepackage[per-mode=symbol,per-symbol =/]{siunitx}
\usepackage[pscoord]{eso-pic}
\usepackage[acronym]{glossaries}
\usepackage{float}
\usepackage{multirow}
\usepackage{array}
\usepackage{setspace}
\usepackage{enumitem}
\usepackage{url}
\usepackage{hhline}
\usepackage[font=footnotesize,labelfont=bf]{caption}
\usepackage{amssymb}
\usepackage{pifont}
\usepackage[hidelinks]{hyperref}
\usepackage{orcidlink}
\usepackage[capitalise]{cleveref}
\usepackage{placeins}
\usepackage{circledsteps}

\newacronym{ac}{AC}{application-class}
\newacronym{amba}{AMBA}{Advanced Microcontroller Bus Architecture}
\newacronym{asic}{ASIC}{application-specific integrated circuit}
\newacronym{ate}{ATE}{automated test equipment}
\newacronym{axi}{AXI4}{Advanced eXtensible Interface 4}
\newacronym{ca}{CA}{command/address}
\newacronym{cmos}{CMOS}{complementary metal-oxide-semiconductor}
\newacronym{cots}{COTS}{commercial off-the-shelf}
\newacronym{d2d}{D2D}{die-to-die}
\newacronym{ddr}{DDR}{double data rate}
\newacronym{dma}{DMA}{direct memory access}
\newacronym{dram}{DRAM}{dynamic random access memory}
\newacronym{dsa}{DSA}{domain-specific architecture}
\newacronym{dsp}{DSP}{digital signal processing}
\newacronym{edram}{eDRAM}{embedded DRAM}
\newacronym{fiwlcsp}{FI-WLCSP}{fan-in wafer-level chip-scale packaging}
\newacronym{fll}{FLL}{frequency-locked loop}
\newacronym{fos}{FOS}{free and open-source}
\newacronym{fpga}{FPGA}{field-programmable gate array}
\newacronym{fsm}{FSM}{finite state machine}
\newacronym{gpos}{GPOS}{general-purpose operating system}
\newacronym{gpt}{GPT}{Globally Unique Identifier Partition Table}
\newacronym{hbm}{HBM}{high-bandwidth memory}
\newacronym{hpc}{HPC}{high performance computing}
\newacronym{hw}{HW}{hardware}
\newacronym{iot}{IoT}{internet of things}
\newacronym{ip}{IP}{intellectual property}
\newacronym{llc}{LLC}{last-level cache}
\newacronym{lpddr}{LPDDR}{low-power double data rate}
\newacronym{mcu}{MCU}{microcontroller unit}
\newacronym{ml}{ML}{machine learning}
\newacronym{nsrrp}{NSRRP}{non-stallable request-response protocol}
\newacronym{ooo}{OoO}{out-of-order}
\newacronym{p2p}{P2P}{point-to-point}
\newacronym{pcb}{PCB}{printed circuit board}
\newacronym{phy}{PHY}{physical interface circuit}
\newacronym{pmca}{PMCA}{programmable many-core accelerator}
\newacronym{ppa}{PPA}{power-performance-area}
\newacronym{rob}{ROB}{reorder buffer}
\newacronym{rpc}{RPC}{reduced pin count}
\newacronym{rpcdram}{RPC DRAM}{reduced-pin-count DRAM}
\newacronym{rtl}{RTL}{register-transfer level}
\newacronym{sbc}{SBC}{single board computer}
\newacronym{sdr}{SDR}{single data rate}
\newacronym[firstplural=static random access memories (SRAMs)]{sram}{SRAM}{static random access memory}
\newacronym{spm}{SPM}{scratchpad memory}
\newacronym[firstplural=systems-on-chip (SoCs)]{soc}{SoC}{system-on-chip}
\newacronym{tdp}{TDP}{thermal design power}
\newacronym{db}{DB}{data bus}
\newacronym{qos}{QoS}{quality of service}

\pgfkeys{/csteps/inner color=white}
\pgfkeys{/csteps/outer color=black}
\pgfkeys{/csteps/fill color=black}
\pgfkeys{/csteps/inner xsep=1pt}
\pgfkeys{/csteps/inner ysep=1pt}

\newcommand{\x}{$\times$}
\renewcommand{\subsubsection}[1]{\paragraph*{\textbf{#1}}}
\newcommand{\rpcdram}{\gls{rpc}~\gls{dram}}

\DeclareSIUnit\GE{GE}
\DeclareSIUnit\bit{b}
\DeclareSIUnit\kGE{\kilo\GE}
\DeclareSIUnit\MGE{\mega\GE}

\def\thetitle{Cheshire: A Lightweight, Linux-Capable \\ RISC-V Host Platform for Domain-Specific Accelerator Plug-In}
\def\thetitleoneline{Cheshire: A Lightweight, Linux-Capable RISC-V Host Platform for Domain-Specific Accelerator Plug-In}

\renewcommand{\baselinestretch}{0.926}

\begin{document}

\title{\thetitle}

\author{Alessandro~Ottaviano\orcidlink{0009-0000-9924-3536},~\IEEEmembership{Student Member, IEEE},
        Thomas~Benz\orcidlink{0000-0002-0326-9676},~\IEEEmembership{Student Member, IEEE},
        \\Paul~Scheffler\orcidlink{0000-0003-4230-1381},~\IEEEmembership{Student Member, IEEE},
        and~Luca~Benini\orcidlink{0000-0001-8068-3806},~\IEEEmembership{Fellow, IEEE}%
        \thanks{A.~Ottaviano, T.~Benz,  and P.~Scheffler  contributed equally to this work.}
        \IEEEcompsocitemizethanks{\IEEEcompsocthanksitem A.~Ottaviano, T.~Benz,  P.~Scheffler and L.~Benini are with the Integrated Systems Laboratory (IIS), ETH Zurich, Switzerland.\protect\\
        E-mail: \{aottaviano,tbenz,paulsc,lbenini\}@iis.ethz.ch
        \IEEEcompsocthanksitem L.~Benini is also with the Department of Electrical, Electronic and Information Engineering (DEI), University of Bologna, Bologna, Italy.\protect
        }%
}

\markboth{IEEE Transactions on Circuits and Systems—II: Express Briefs (TCAS-II)}%
{Ottaviano \MakeLowercase{\textit{et al.}}: \thetitleoneline}

\maketitle

\ifx\showrebuttal\undefined
    \newcommand{\revision}[1]{{#1}}
\else
    \newcommand{\revision}[1]{{\textcolor{blue}{#1}}}
\fi

\ifx\showrevision\undefined
    \newcommand{\todo}[1]{{#1}}
\else
    \newcommand{\todo}[1]{{\textcolor{red}{#1}}}
    \AddToShipoutPictureFG{%
        \put(%
            8mm,%
            \paperheight-1.5cm%
            ){\vtop{{\null}\makebox[0pt][c]{%
                \rotatebox[origin=c]{90}{%
                    \huge\textcolor{red!75}{\reviewpass}%
                }%
            }}%
        }%
    }
    \AddToShipoutPictureFG{%
        \put(%
            \paperwidth-6mm,%
            \paperheight-1cm%
            ){\vtop{{\null}\makebox[0pt][c]{%
                \rotatebox[origin=c]{90}{%
                    \huge\textcolor{red!30}{ETH Zurich - Unpublished - Confidential - Draft - Copyright Thomas, Ale, Paul 2023}%
                }%
            }}%
        }%
    }
\fi

\begin{abstract}
Power and cost constraints in the internet-of-things (IoT) extreme-edge and TinyML domains, coupled with increasing performance requirements, motivate a trend toward heterogeneous architectures. 
These designs use energy-efficient application-class host processors to coordinate compute-specialized multicore accelerators, amortizing the architectural costs of operating system support and external communication. %
This brief presents \emph{Cheshire}, a lightweight and modular 64-bit Linux-capable host platform designed for the seamless plug-in of domain-specific accelerators.
It features a unique low-pin-count DRAM interface, a last-level cache configurable as scratchpad memory, and a DMA engine enabling efficient data movement to or from accelerators or DRAM. 
It also provides numerous optional IO peripherals including UART, SPI, I2C, VGA, and GPIOs.
Cheshire's synthesizable RTL description, comprising all of its peripherals and its fully digital DRAM interface, is available free and open-source.
We implemented and fabricated Cheshire as a silicon demonstrator called \emph{Neo} in TSMC's 65nm CMOS technology.
At 1.2\,V, Neo achieves clock frequencies of up to 325\,MHz while not exceeding 300\,mW in total power on data-intensive computational workloads.
Its RPC DRAM interface consumes only 250\,pJ/B and incurs only 3.5\,kGE in area for its PHY while attaining a peak transfer rate of 750\,MB/s at 200\,MHz.
\end{abstract}


\section{Introduction}

With Koomey's law\footnote{Observed from the 1940s to early 2000s, Koomey's Law states that the energy efficiency of computers doubles every 1.57 years.}~\cite{KOOMEY} slowing down, computer architects turn to hardware specialization to meet the rising energy efficiency requirements of data-intensive applications like \gls{ml} and near-sensor processing.
Today's heterogeneous architectures couple conventional \gls{ac} host processors to \glspl{dsa} or \glspl{pmca}~\cite{DSC_I, PATTERSON_HENNESSY_TURING, 9748063, 9763876}, improving energy efficiency while maintaining high programmability and \gls{gpos} support.
Thus, to maximize compute efficiency in heterogeneous systems, architects should maximize the area and energy resources spent on efficient accelerators and minimize the cost and energy footprint of \gls{ac} hosts.

\revision{H}eterogeneous architectures are established in mobile and high-performance applications\revision{: silicon-proven industrial~\cite{pi0, Jetson, unmatched} and academic~\cite{ESP, BYOC} \glspl{soc} typically target large power envelopes or multiple \gls{ac} cores%
.}
\revision{However, heterogeneity is} now penetrating the \gls{iot} extreme-edge and TinyML domains where power and cost constraints are extremely tight. Heterogeneous \glspl{soc} in these domains~\cite{KRAKEN, VEGA} typically \revision{use} lightweight 32-bit microcontrollers without \gls{ac} capabilities or \gls{gpos} support as management processors.
The recent \emph{HULK-V}~\cite{HULKV_DATE} \gls{soc} demonstrates a more capable host by coupling a Linux-capable 64-bit {RISC-V} manager core to an eight-core \gls{pmca} for high-end tinyML applications. However, it has not been silicon-proven, and its \gls{pmca}'s ad-hoc integration does not provide a reusable interface. Furthermore, {HULK-V}'s external HyperRAM~\cite{HYPER} memory interface is low-bandwidth, limiting scalability and potentially incurring host-accelerator memory bottlenecks.

To address these shortcomings, we present \emph{Cheshire}, an energy-efficient, Linux-capable, 64-bit {RISC-V} host platform for \revision{the seamless heterogeneous plug-in of \glspl{dsa} such as \cite{pulpnn1, manor1}}. 
Cheshire is fully modular and provides a configurable interconnect, numerous optional peripherals, and a \gls{dma} engine to decouple host-\gls{dsa} communication. 
It integrates the first synthesizable open-source interface for \glsunset{dram}{\rpcdram}, a recent low-pin-count \gls{dram} solution incurring lower integration cost and effort than \gls{lpddr}~\gls{dram}, but significantly higher bandwidth than HyperRAM at comparable energy efficiency.
Cheshire's synthesizable \gls{rtl} description \revision{and} \glsunset{fpga}\gls{fpga} implementation are available \revision{open-source}\footnote{\url{https://github.com/pulp-platform/cheshire} for Cheshire and \url{https://github.com/pulp-platform/rpc_dram_controller} for its RPC DRAM interface.}.
We present the following contributions:
\begin{itemize}
    \item A minimal, customizable, and energy-efficient Linux-capable RV64GC host platform that can easily be co-integrated with on-chip or chiplet \glspl{dsa}; we provide a configurable \gls{axi} interconnect and a digital \gls{d2d} interface.
    \item The first \emph{fully digital}, technology-independent \gls{rpc}-\gls{dram}-compliant memory interface, which incurs only 22 switching IOs and \SI{3.5}{\kGE} in PHY area. It enables memory accesses at only \SI{250}{\pico\joule\per\byte}  while attaining a peak transfer rate of \SI{750}{\mega\byte\per\second} at \SI{200}{\mega\hertz}, outperforming existing HyperRAM solutions. \revision{To the best of our knowledge, this is the first characterization of \gls{rpcdram} in a taped-out open-source \gls{soc}.}
    \item An agile memory system, enabling accesses to external \gls{rpc} \gls{dram} in only 8 cycles for a \SI{32}{\byte} transfer while interacting with \gls{dsa} memory space only when desired.
    \item A standalone demonstrator chip called \textit{Neo}, fabricated in TSMC's \SI{65}{\nano\metre} node. At \SI{1.2}{\volt}, Neo achieves clock frequencies of up to \SI{325}{\mega\hertz} while consuming less than \SI{300}{\milli\watt} during data-intensive computational workloads.
\end{itemize}

\section{Architecture}\label{sec:architecture}

In the following section, we present the overall architecture of Cheshire, followed by an in-depth discussion of \revision{the} fully-digital, technology-agnostic {\rpcdram} interface.

\subsection{Cheshire Platform}
\label{subsec:soc-arch}

\begin{figure}[t]
    \centering
    \includegraphics[width=\columnwidth]{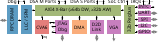}
    \caption{%
        Architecture of the Cheshire platform. The \gls{axi} crossbar provides a configurable number of \textbf{M}anager and \textbf{S}ubordinate ports toward a \gls{dsa}. 
    }
    \label{fig:soc-arch}
\end{figure}

Cheshire is based on the energy-efficient 64-bit CVA6~\cite{zaruba2019cost} \gls{ac} processor. It includes all hardware necessary to boot and run a \gls{gpos} like Linux \revision{autonomously}, such as RISC-V-compliant \revision{core-local and platform} interrupt controllers, various optional \revision{standard} IO interfaces to access external storage and peripherals, and a \gls{dram} interface. \revision{Off-chip \gls{dram} is essential as the $8$-\SI{16}{\mega\byte} memory footprint of simple embedded Linux systems \cite{opdenacker2017linux} usually does not fit into on-chip memory.}

\Cref{fig:soc-arch} shows Cheshire's architecture. A crossbar~\cite{AXIPULP} using \revision{Arm's} widespread \gls{axi} protocol\revision{\cite{arm2021amba}} connects the CVA6 processor to {\rpcdram}, the \gls{dsa}, and other Cheshire-internal components. The crossbar's address width, data width, and the number of \revision{\gls{axi}} \gls{dsa} manager and subordinate ports are configurable to suit the target system's bandwidth and addressing needs. Simpler subordinates without burst or out-of-order transaction support are attached through a lightweight, extensible Regbus\cite{regbus} demultiplexer, minimizing the crossbar's area and energy footprint.

Cheshire's {\rpcdram} is connected through a configurable \gls{llc}. Each of the \gls{llc}'s ways may individually be configured to serve as a \gls{spm} at runtime, providing the host with fast internal \gls{sram} when needed. 
A RISC-V compliant debug module, backed by a JTAG transport module, enables live external debugging of CVA6 and any configured number of external {RISC-V} harts, e.g., those in \gls{pmca}~\glspl{dsa}.
Likewise, the interrupt controllers support a configurable number of external sources and targets.

Cheshire provides various optional peripherals, including a \revision{flexible \gls{axi}}~\gls{dma} engine\revision{\cite{benz2023highperformance}} for efficient data movement, a VGA controller for display output, and a digital \gls{d2d} link for communication with off-chip systems. It also provides a UART for serial communication, a GPIO module, and I2C and SPI hosts to access external peripherals. \revision{All peripherals seamlessly integrate through \gls{axi} or Regbus interfaces and provide well-established feature sets for full compatibility with existing Linux drivers.}  An additional \emph{\gls{soc} control} port connects to Cheshire-external on-chip devices essential for operation, such as clock generators, IO multiplexers, or clock and power domain controllers.

Cheshire has a built-in boot ROM, allowing for passive preloading through JTAG, UART, or the \gls{d2d} link or autonomous boot from an external SPI Flash, I2C EEPROM, or SD card with \gls{gpt} support. If configured, it invokes an external ROM first to set up the surrounding chip through the \gls{soc} control port. \revision{Compiled with \texttt{-Os} flags and full-program link-time optimization, Cheshire's boot ROM is \SI{7.2}{\kibi\byte} in size.}

Although we use a single CVA6 as Cheshire's \gls{ac} processor, multiple coherent CVA6 cores or another \gls{ac} processor could be integrated instead; here, we focus on a minimal \gls{ac} host.

We provide an open-source \gls{fpga} implementation of Cheshire, currently targeting Digilent's Genesys II board\revision{, which} enables the rapid prototyping and characterization of heterogeneous systems.

\subsection{RPC DRAM Interface}
\label{subsec:rpc_uarch}

\begin{figure}[t]
    \centering
    \includegraphics[width=0.95\columnwidth]{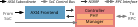}
    \caption{%
        Architecture of the {\rpcdram} memory interface. %
    }
    \label{fig:rpc_arch}
\end{figure}

\subsubsection{\revision{Background}}\label{subsec:rpc_protocol}

\revision{\gls{rpcdram} is specified to use a minimum number of signals to deliver \glsunset{ddr}\gls{ddr}3-level in-system bandwidth~\cite{RPC-spec}.
Its defining feature is its interface, which multiplexes data, addresses, and commands onto a common data bus (\emph{DB})\glsunset{db} to minimize the needed pins.
\Gls{ca} control can be sent concurrently with data by using a single multi-function pin to transmit serial commands, offering the high transaction efficiency of \gls{ddr}3 at the cost of just 22 switching signals for a 16-bit wide \emph{DB}. }
\revision{The strobe pins \emph{DQS} and \emph{DQS\#} sample \emph{DB} with a preamble identical to \gls{ddr}3}.

\revision{
Alternative \gls{dram} solutions with low-pin-count interfaces have been proposed.}
\revision{Cypress' \emph{HyperRAM}~\cite{HYPER} requires only 12 switching IOs for an 8-bit shared bus. However, transfer rates are limited to \SI{400}{\mega\byte\per\second} at \SI{200}{\mega\hertz} or less, and its self-refresh precludes advanced controller-side scheduling.
}
\revision{
Antmicro proposes an FPGA-based Rowhammer testing platform based on the LiteDRAM memory controller~\cite{LITEDRAM}.
The controller supports several \gls{ddr} memory types and recently added an %
\gls{rpc} \gls{phy} implementation. Although open-source%
, this \gls{phy} targets \glspl{fpga} only.
}

\subsubsection{\revision{Interface}} \revision{\Cref{fig:rpc_arch} depicts our} {\rpcdram} interface. \revision{It is comprised of} two parts: a \textit{controller} implementing the off-chip \gls{rpc} protocol~\cite{RPC-spec} and \revision{an \gls{axi}} \textit{frontend} implementing an \gls{axi}-compliant subordinate. %
\revision{To enable easy adaptation to on-chip protocols other than \gls{axi}, the controller and frontend are connected through a generic interface we call \glsunset{nsrrp}\emph{non-stallable request-response protocol (NSRRP)}; its datawidth is \SI{256}{\bit} or one \emph{word} in the \rpcdram~standard. In the following, we will discuss the controller and \gls{axi} frontend in detail.}

\subsubsection{Controller}
\label{subsec:rpc_ctrl}

\begin{figure}[t]
    \centering
    \includegraphics[width=\columnwidth]{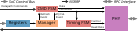}
    \caption{Architecture of the {\rpcdram} Controller.}
    \label{fig:rpc_ctrl}
\end{figure}

\Cref{fig:rpc_ctrl} depicts the internal controller architecture. %
\revision{As shown, the controller receives \emph{datapath commands} from the frontend. These generic commands} are passed to the \emph{command \glsunset{fsm}\gls{fsm}}, which decomposes \revision{them} into {\rpcdram}-\revision{specific} commands. %
For example, a \revision{generic} datapath read is decomposed into \revision{1)} an \emph{activate} of the corresponding bank and row, \revision{2)} a \emph{read} of $N$ \revision{consecutive \rpcdram~words}, and \revision{3)} a \emph{precharge} to close the bank and prepare it for the next access. %

In addition to datapath commands \revision{issued by the frontend}, the command \gls{fsm} also handles \emph{management commands} \revision{issued} by \revision{a} \emph{manager} module \revision{inside the controller}.
The manager has three responsibilities:
\revision{1)} it \emph{initializes} the {\rpcdram} \revision{device} on startup, %
\revision{2)} it periodically \emph{refresh\revision{es}} active banks,  and %
\revision{3)} it \revision{performs} \emph{ZQ calibration} when necessary. %
\revision{For these tasks, the manager uses configurable timing parameters, which can be set through a memory-mapped register file.}

The \revision{command FSM passes its generated} {\rpcdram} commands to the \emph{timing \gls{fsm}}, which performs two tasks:
\revision{1)} it \emph{times commands}, ensuring that \revision{they adhere} to protocol constraints like \revision{cycle alignment and minimum delays}, and %
\revision{2)} it \emph{times the \revision{physical} interface}, which includes controlling the chip select signals, \revision{gating} the output strobe, and multiplexing data, mask, and commands \revision{onto the \emph{DB}}.

\begin{figure}[t]
    \centering
    \includegraphics[width=0.9\columnwidth]{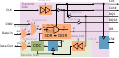}
    \caption{\revision{Architecture of the \gls{rpc} \gls{phy}.}}
    \label{fig:rpc_phy}
\end{figure}

The \emph{physical interface circuit} (PHY), shown in \Cref{fig:rpc_phy}, %
implements a low-power, digital-only, technology-agnostic {\rpcdram} physical layer without internal clock generation. %
\revision{As shown, the PHY is composed of two parts: the \emph{transmit side} sends data to the \gls{rpcdram}, and the \emph{receive side} accepts data from the \gls{rpcdram}}.

The \textit{transmit} side creates 90- and 270-degree phase-shifted clocks \revision{with} a configurable delay line to drive the strobe signals \revision{DQS and DQS\#}, which are selectively enabled by the timing \gls{fsm}. %
Any \SI{256}{\bit} \revision{data} word\revision{s} \revision{to be sent to the {\rpcdram} are first serialized} to \SI{32}{\bit} \emph{subwords}; the timing \gls{fsm} then arbitrates between sending \SI{32}{\bit} commands and \revision{\SI{32}{\bit}} data subwords using a \revision{multiplexer}. %
\revision{The chosen payload} is converted from \gls{sdr} to \gls{ddr} \revision{using clock-driven multiplexers as shown} and placed on the \revision{\emph{DB}}. %

The \textit{receive} side accepts the strobe signal \emph{DQS} in phase with the data read from the {\rpcdram} chip through the \emph{DB} pins. %
The received data is converted to \gls{sdr}, then sampled by a delayed strobe generated with another configurable delay line. 
After passing through a clock domain crossing, the read data is packed to form complete \SI{256}{\bit} words and then sent back to the \gls{axi} frontend over a\revision{n} \gls{nsrrp} channel.

\subsubsection{\revision{AXI4} Frontend}
\label{sec:rpc_axi_intf}

\begin{figure}[t]
    \centering
    \includegraphics[width=\columnwidth]{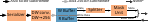}
    \caption{Architecture of \revision{the} \gls{rpc} interface's \gls{axi} frontend.}
    \label{fig:rpc_axi}
\end{figure}

The frontend, shown in  \Cref{fig:rpc_axi}, implements an \gls{axi}-compliant subordinate.
Incoming requests are first \emph{serialized} as the {\rpcdram} controller operates strictly in order. \revision{In the current design}, transfers from different \revision{\gls{axi}} IDs are handled first come, first serve; \revision{we plan to implement} transfer prioritization using \gls{axi}'s \gls{qos} \revision{signals in} future versions.  %
After serialization, a \emph{datawidth converter} converts \revision{the {\rpcdram} interfaces' configured} datawidth \revision{(\SI{64}{\bit} in the case of Neo)} to \gls{rpc}'s \SI{256}{\bit} word size. 

While \gls{axi} \revision{allows} transfers \revision{to} be stalled on any beat, \gls{rpc}~bursts cannot be stalled once launched. %
Hence, both reads and writes are \emph{buffered}. 
\emph{Write} data is buffered and released once the buffer contains all data needed for the next write. %
\emph{Read} data is forwarded to the \gls{axi} bus as soon as possible to minimize latency and buffered only on \gls{axi} bus stalls. %

\revision{The} \emph{splitter} splits \gls{nsrrp} transactions \revision{at} \SI{2}{\kibi\byte} boundaries to comply with the \gls{rpc} protocol. %
{\rpcdram} implements unaligned transfers by introducing a \emph{first} and a \emph{last} \emph{mask}. %
The masks are transmitted between the write command and the corresponding data. %
\revision{The} \emph{mask unit} derives these two masks from the strobe information included in \gls{axi} requests. %

\section{Evaluation}
\label{sec:evaluation}

We evaluate Cheshire through a silicon demonstrator named \emph{Neo}, which was designed, implemented, and fabricated using {TSMC}'s \SI{65}{\nano\metre} node with nine metal layers. %
Neo was synthesized using {Synopsys} {Design} {Compiler} 2019.03 and implemented using {Cadence} {Innovus} {2019.10}, targeting a \SI{200}{\mega\hertz} system clock in the SS corner at \SI{125}{\celsius} and using low-$V_T$ cells. %
We will first describe how we configured Cheshire for Neo, then characterize the {\rpcdram} and \revision{the} \gls{axi} \gls{dsa} interfaces in terms of functional performance (\cref{subsec:eval:functional}) and silicon implementation performance (\cref{subsec:eval:silicon}). 

\subsection{Silicon Demonstrator}
\label{sec:eval:silicon}

\begin{figure}[t]
    \centering
    \includegraphics[width=0.9\columnwidth]{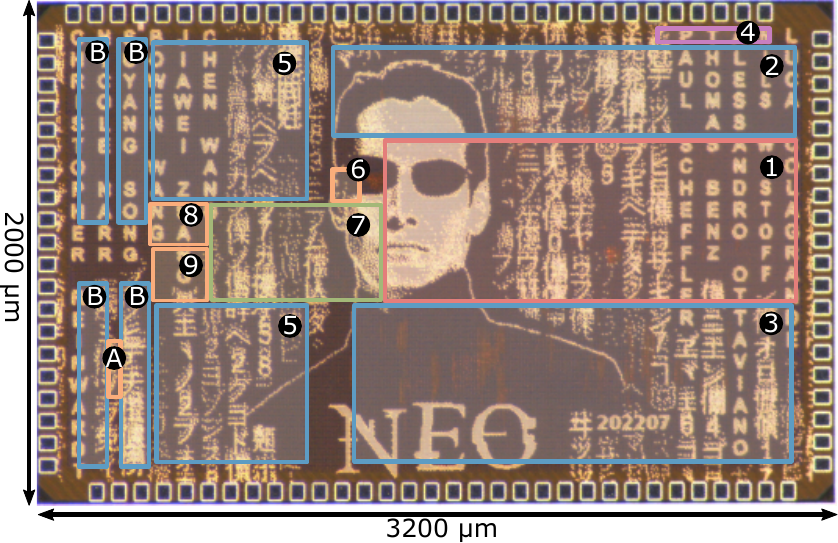} %
    \caption{%
        Die shot of \emph{Neo}: %
        \Circled{\textbf{1}}~CVA6 and uncore region, %
        \Circled{\textbf{2}}~D-cache, %
        \Circled{\textbf{3}}~I-cache, %
        \Circled{\textbf{4}}~\gls{d2d} link, %
        \Circled{\textbf{5}}~\gls{spm}, %
        \Circled{\textbf{6}}~FLL, %
        \Circled{\textbf{7}}~\gls{axi} interconnect and DMA, %
        \Circled{\textbf{8}}~\gls{rpc} manager, %
        \Circled{\textbf{9}}~\gls{rpc} \gls{axi} interface, %
        \Circled{\textbf{A}}~\gls{rpc} controller, %
        \Circled{\textbf{B}}~\gls{rpc} buffer. %
    }
    \label{fig:neo-die}
\end{figure}

\Cref{fig:neo-die} shows an annotated die shot of Neo. Its \SI{6.4}{\square\milli\metre} die is housed in a {QFN64} package. %
We use a core voltage of \SI{1.2}{\volt} and a global {IO} voltage of \SI{1.5}{\volt} to comply with the {\rpcdram} interface.
We configure Cheshire without \gls{dsa} ports as we do not integrate any accelerator in this demonstrator.
Neo features \SI{128}{\kibi\byte} of \gls{spm}, \SI{64}{\bit} data, and \SI{48}{\bit} addresses; its CVA6 core is configured with \SI{32}{\kibi\byte} 8-way level-one data and instruction caches.
The {\rpcdram} frontend is configured with \SI{8}{\kibi\byte} buffers for read and write each.
We include an on-chip \gls{fll} for internal clock generation and control.

\begin{figure}[t]
    \centering
        \subfloat{\includegraphics[height=3.1cm]{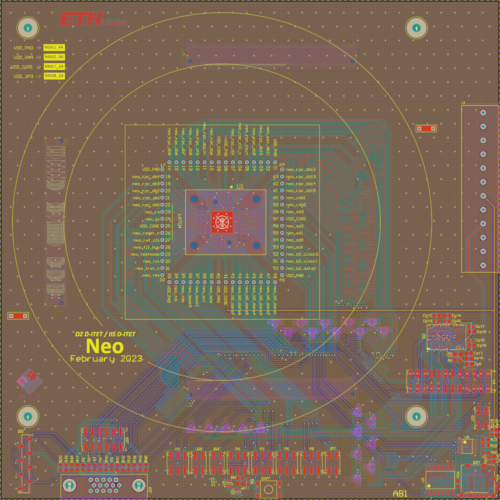}%
        \label{fig:neo-board-concept}%
        }
        \qquad
        \subfloat{\includegraphics[height=3.1cm]{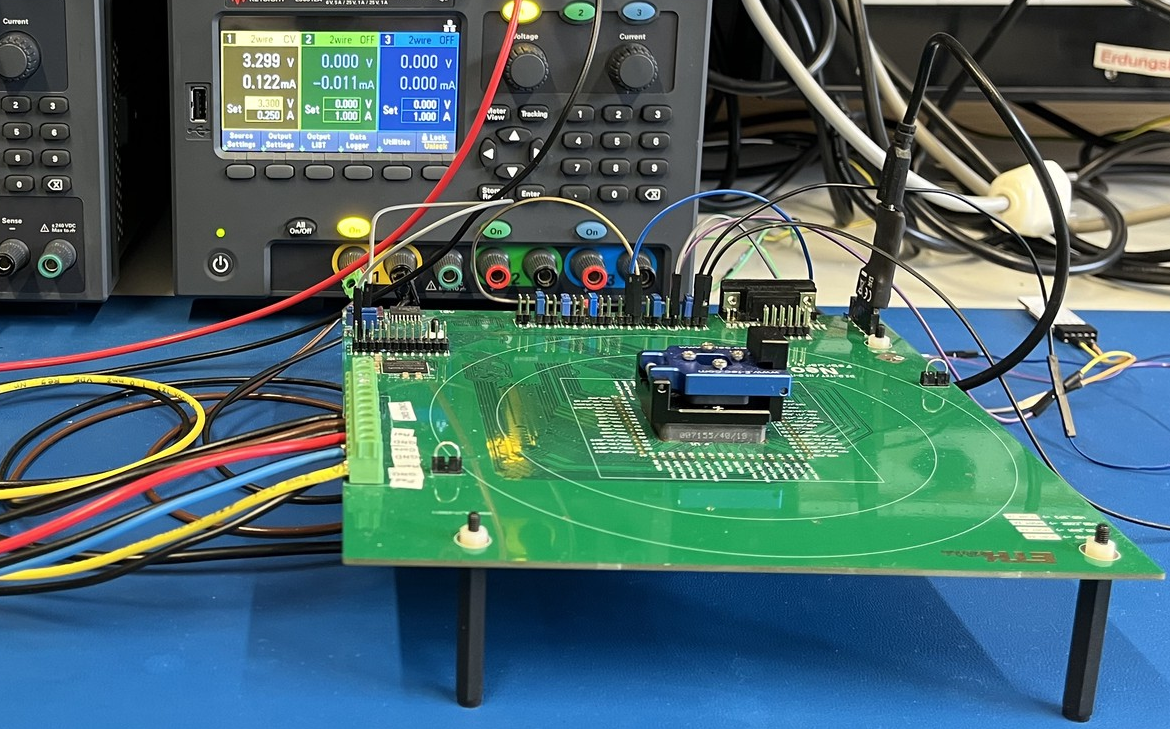}%
        \label{fig:neo-board-fab}%
        }
    \caption{%
        Evaluation PCB to characterize \emph{Neo} standalone and using industry-grade \gls{ate} equipment. %
        (left) Layout of the board (right) fabricated PCB. %
        \revision{\gls{dram} traces are \SI{25.23}{\centi\metre} long, \SI{0.15}{\milli\metre} wide with {JEDEC}-compliant single-ended impedance of \SI{50}{\ohm} and differential of \SI{90}{\ohm}.}
    }
    \label{fig:neo-board}
\end{figure}

We test Neo on a custom-made bring-up {PCB} shown in \Cref{fig:neo-board}. This board can be connected to an industry-grade \gls{ate} device (Advantest~SoCV93000) or operated standalone. %
In standalone mode, the board provides Neo with a \SI{32}{\kilo\hertz} reference clock to lock its internal \gls{fll} and a debounced reset; %
it also features a set of peripherals and connectors to boot and run applications, \revision{including a \SI{32}{\mebi\byte} \rpcdram~chip (EM6GA16LBXA-12H) connected to Neo and used for testing and measurements}. %
User interaction may happen through {UART} and {VGA} output. %

\subsection{Functional Performance}
\label{subsec:eval:functional}

Functional evaluation is performed through cycle-accurate \revision{RTL} simulation. For all presented results, we leverage the efficient data movement capabilities of the \gls{dma} engine, which enables decoupled, high throughput host-\gls{dsa} transfers and frees CVA6 from handling data movement tasks.

\begin{figure}[t]
    \centering
    \includegraphics[width=\columnwidth]{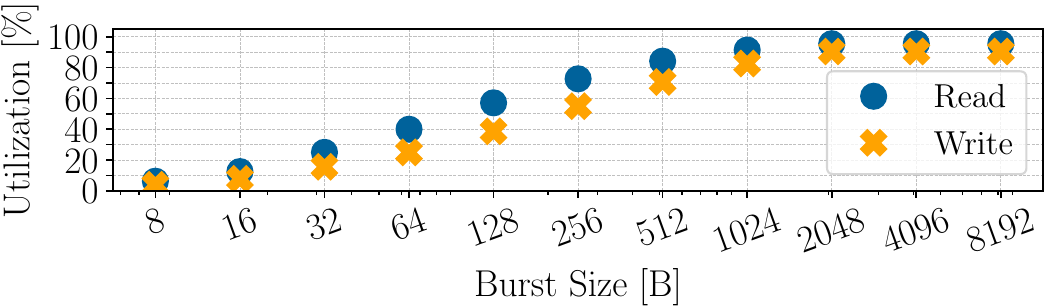}
    \caption{%
        Relative {\rpcdram} bus utilization on reads and writes. %
    }
    \label{fig:neo-rpc-func}
\end{figure}

{\rpcdram} specifies a maximum bus clock frequency of \SI{933}{\mega\hertz}, outperforming competing solutions. Provided Neo's \SI{16}{\bit} \emph{DB} bus and an operating frequency of \SI{200}{\mega\hertz} with DDR signaling, the peak attainable throughput is $\Theta$ = $ \alpha\cdot\text{\SI{800}{\mega\byte\per\second}}$, where $\alpha$ is the relative bus utilization. 
\Cref{fig:neo-rpc-func} shows the attained \gls{rpc} bus utilization for both the read and write datapaths. 
The \gls{dma} is programmed to issue write and read transfers at increasing burst sizes starting from \SI{8}{\byte}.
The interface bus utilization plateaus close to peak utilization ($\alpha = 1$) for bursts \SI{2}{\kibi\byte} in size or larger as they are decomposed into smaller transfers by the \gls{axi} frontend's transfer splitter.
While the \gls{rpc} pre- and postamble incur a constant, protocol-defined overhead independent of transfer size and controller implementation, bus utilization on reads is on average 1.3\x~higher than on corresponding writes; this is because read data is passed to the \gls{axi} bus as soon as possible whereas writes are deferred until enough data is buffered.

{\rpcdram} proves its superiority in  attainable bandwidth and area footprint against one of its closest competitors, HyperBus, which was implemented in several academic works~\cite{VEGA, HULKV_DATE}.
HyperRAM occupies more \glsunset{pcb}\gls{pcb} area, has a much lower maximum data rate of \SI{400}{\mega\byte\per\second}, and is limited to a maximum frequency of only \SI{200}{\mega\hertz}~\cite{HYPER}, restricting its applicability in high-bandwidth energy-efficient scenarios.

\subsection{Silicon performance}\label{subsec:eval:silicon}

\subsubsection{Area breakdown}\label{subsubsec:eval:silicon:area}

\begin{figure}[t]
    \centering
    \includegraphics[width=\linewidth]{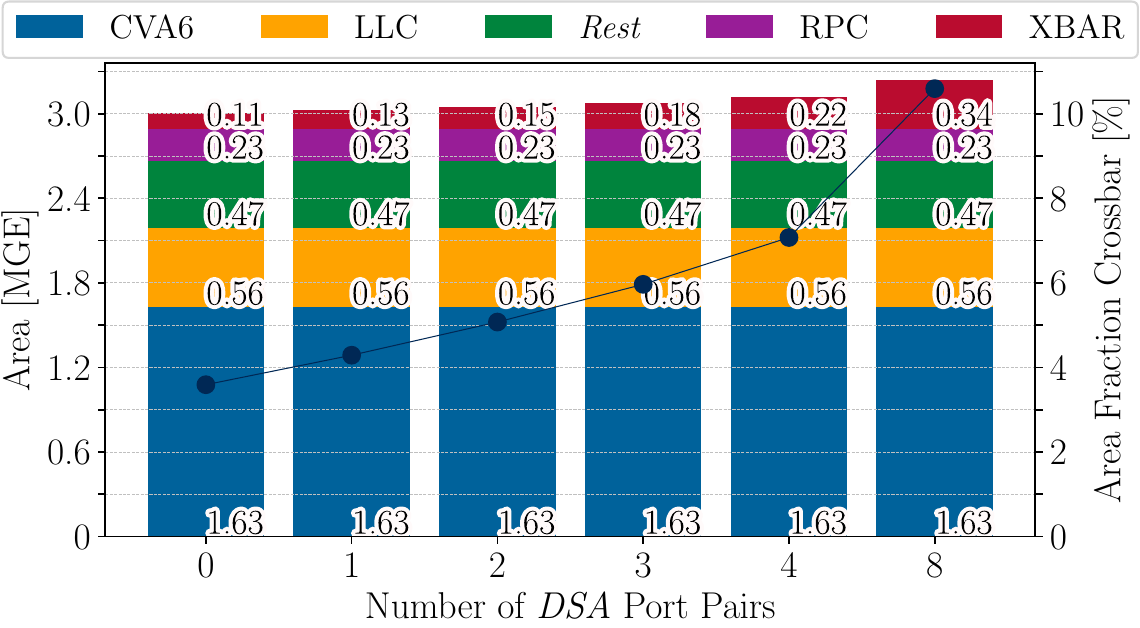}
    \caption{%
        Area breakdown of Cheshire implemented in TSMC65 and relative contribution of the crossbar for different numbers of \gls{dsa} port pairs.
        \emph{Rest} denotes the area of the DMA, peripherals, and interconnect adapters. %
    }
    \label{fig:neo-area}
\end{figure}

\Cref{fig:neo-die} shows Neo's die, highlighting the approximate location and size of its main components.
\Cref{fig:neo-area} reports Cheshire's exact area breakdown in \si{\kGE}, progressively increasing the number of \gls{dsa} manager-subordinate port pairs attached to its main \gls{axi} crossbar. 
The leftmost bar corresponds to Neo's configuration without \gls{dsa} ports.
In all considered cases, CVA6 dominates Cheshire's area, while the {\rpcdram} controller accounts for at most \SI{7.6}{\percent}.
As we increase the number of \gls{dsa} ports, \revision{the} all-to-all \gls{axi} crossbar grows from \SI{3.6}{\percent} to \SI{10.6}{\percent} of Cheshire, increasing its area by at most \SI{7.8}{\percent} for eight port pairs compared to Neo's configuration without \gls{dsa} ports. While \revision{crossbar} scaling is a limiting factor for large numbers of port pairs, \glspl{dsa} with many \revision{managers} or subordinates can use a sub-interconnect, larger data widths, or sparse connectivity to facilitate scaling.

\begin{figure}[t]
    \centering
    \includegraphics[width=\linewidth]{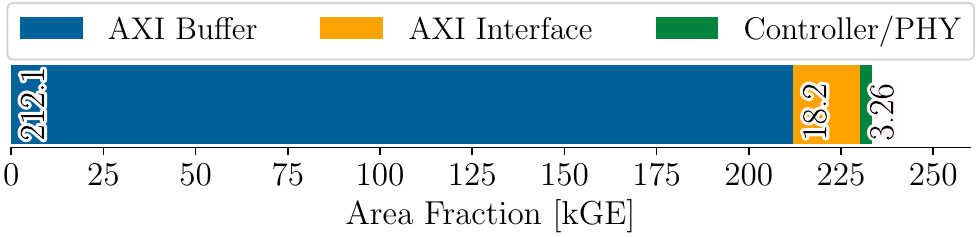}
    \caption{%
        Area breakdown of the {\rpcdram} interface. %
        When configured as in Neo, the \emph{\gls{axi} buffer} and the \emph{\gls{axi} Interface} occupy most of the area. %
    }
    \label{fig:rpc-area}
\end{figure}

\Cref{fig:rpc-area} shows the area breakdown of Neo's {\rpcdram} controller.
The manager module, command/timing \gls{fsm}, and digital \gls{phy} occupy only \SI{1}{\percent} or \SI{3.5}{\kGE} of the controller's area, confirming the extremely low area overhead incurred by the \gls{rpc} protocol for its memory interface.
Most of the controller's area overhead is due to the buffers holding \gls{axi} beats. In Neo, these buffers are over-provisioned to simplify the initial design of \revision{the} \gls{axi} frontend, but their size can be further reduced in future versions. \revision{Despite this, our controller occupies only \SI{6.3}{\percent} of the area and \SI{33}{\percent} of the beachfront of an existing \SI{65}{\nano\meter} full-pin-count \gls{ddr}3 controller \cite{TUK}.}

\subsubsection{Energy efficiency}\label{subsubsec:eval:silicon:energy}
\begin{figure}
    \centering
    \includegraphics[width=\columnwidth]{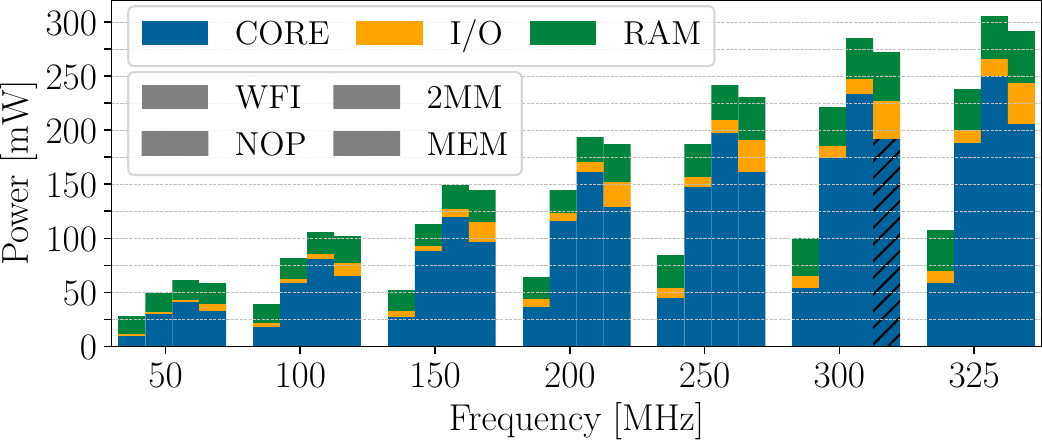}
    \caption{Power consumption of {Neo} for the four workloads: \emph{WFI}, \emph{NOP}, \emph{2MM}, and \emph{MEM}. The power is split into the three power domains of {Neo}.}
    \label{fig:neo-power}
\end{figure}

\Cref{fig:neo-power} shows Neo's power consumption for different scenarios and frequencies as measured on \revision{the} bring-up board; each scenario explores an operational corner with different computational and memory intensities. \revision{We focus here on evaluating the Cheshire platform; for further benchmarks of the CVA6 processor, we refer the reader to its publication \cite{zaruba2019cost}.}
\revision{The} bring-up board provides three power domains with their supplies: \textsc{core}, \textsc{io}, and \textsc{ram}. %
\textsc{core} feeds Neo's core area logic and \glspl{sram}, \textsc{io} provides power to its pads, and \textsc{ram} supplies the {\rpcdram} memory chips. %

In the \emph{WFI} scenario, CVA6 \revision{is waiting for an interrupt}, idling \revision{without} fetching or decoding instructions\revision{; this provides a power baseline with minimal switching}. In \emph{NOP}, CVA6 loops on a body of \texttt{nop}s, establishing a floor for actively fetching, branching, and decoding workloads with few stalls.
\emph{2MM} \revision{runs an optimized} double-precision floating-point \revision{matrix} multiplication \revision{with arguments and results in \rpcdram, keeping reusable matrix tiles in \gls{spm}}. \revision{\emph{MEM} writes high-throughput bursts to \rpcdram~using the \gls{dma} engine.}

At a \textsc{core} supply of \SI{1.2}{\volt}, Neo achieves clock frequencies of up to \SI{325}{\mega\hertz} and remains within a \SI{300}{\milli\watt} power envelope even in data-intensive computational scenarios like \emph{2MM}. All power contributions scale linearly with frequency as expected, and \textsc{core} power dominates in almost all cases; at \SI{200}{\mega\hertz}, \SI{69}{\percent} of \emph{MEM} power is consumed in \textsc{core}.
Since \revision{the} version of {\rpcdram} interface \revision{proposed in this manuscript} does not make use of the technology's \emph{Deep Power Down} state, all benchmarks show an \textsc{ram} idle power consumption. 
\revision{That being said, the {\rpcdram} interface's \textsc{io} power at \SI{200}{\mega\hertz} for \textit{MEM} is \SI{45}{\percent} lower than that of an existing \SI{65}{\nano\meter} \gls{ddr}3 interface under high load \cite{TUK}}.

We use \emph{MEM} to compute {\rpcdram}'s transfer efficiency. We consider only the write direction, representing the worst-case scenario regarding buffering as discussed in \cref{sec:architecture}. 
Given the maximum bandwidth measured in \cref{subsec:eval:functional}, {\rpcdram}'s interface energy per transferred byte is $\Gamma = \frac{P_{tot}}{\Theta} \simeq \text{\SI{250}{\pico\joule\per\byte}}$. This result is comparable to reported energy-per-byte consumptions in recent works integrating lower-bandwidth memories like HyperRAM~\cite{VEGA}.

\section{Conclusion}
\label{sec:conclusions}
This work presents Cheshire, a lightweight, modular, and open-source 64-bit Linux-capable host platform designed \revision{to plug in \glspl{dsa} seamlessly.}
Cheshire features a unique {\rpcdram} interface, a last-level cache, configurable as scratchpad memory, and a DMA engine enabling efficient data movement to or from accelerators or \gls{dram}. 
It also provides numerous optional peripherals, including UART, SPI, I2C, a GPIO module, and VGA for display output.
We implement and fabricate Cheshire as a silicon demonstrator called \emph{Neo} in TSMC's 65nm CMOS technology.
At \SI{1.2}{\volt}, Neo achieves clock frequencies of up to \SI{325}{\mega\hertz} while not exceeding \SI{300}{\milli\watt} in total power even in data-intensive computational scenarios.
Its \gls{dram} interface consumes only \SI{250}{\pico\joule\per\byte} while incurring only \SI{3.5}{\kGE} in area for its \gls{phy} and attaining a peak transfer rate of \SI{750}{\mega\byte\per\second} at \SI{200}{\mega\hertz}.

\section*{Acknowledgments}
\revision{We thank the reviewers for their feedback.}
We thank %
C. Jinfan, %
V. Krishna, %
R. Zhou, %
N. Narr, %
C. Reinwardt, %
S. Song, %
B. Wang, %
C. Wang, %
J. Zhang, %
N. Wistoff, and %
L. Colagrande
for their contributions to this research. %
We also thank {Etron Technology Inc.}, in particular %
R. Crisp,
for their support. %
This work is supported in part through the FRACTAL (877056) project, which received funding from the ECSEL Joint Undertaking (JU) and TRISTAN (101095947) project, which received funding from the HORIZON KDT-JU programme.%

\newpage
\renewcommand{\baselinestretch}{1.0}

%

%

%

\begin{thebibliography}{10}
\providecommand{\url}[1]{#1}
\csname url@samestyle\endcsname
\providecommand{\newblock}{\relax}
\providecommand{\bibinfo}[2]{#2}
\providecommand{\BIBentrySTDinterwordspacing}{\spaceskip=0pt\relax}
\providecommand{\BIBentryALTinterwordstretchfactor}{4}
\providecommand{\BIBentryALTinterwordspacing}{\spaceskip=\fontdimen2\font plus
\BIBentryALTinterwordstretchfactor\fontdimen3\font minus
  \fontdimen4\font\relax}
\providecommand{\BIBforeignlanguage}[2]{{%
\expandafter\ifx\csname l@#1\endcsname\relax
\typeout{** WARNING: IEEEtran.bst: No hyphenation pattern has been}%
\typeout{** loaded for the language `#1'. Using the pattern for}%
\typeout{** the default language instead.}%
\else
\language=\csname l@#1\endcsname
\fi
#2}}
\providecommand{\BIBdecl}{\relax}
\BIBdecl

\bibitem{KOOMEY}
J.~Koomey, S.~Berard, M.~Sanchez, and H.~Wong, ``Implications of historical
  trends in the electrical efficiency of computing,'' \emph{IEEE Ann. Hist.
  Comput.}, vol.~33, no.~3, pp. 46--54, 2011.

\bibitem{DSC_I}
Y.~Chi, W.~Qiao, A.~Sohrabizadeh, J.~Wang, and J.~Cong, ``Democratizing
  domain-specific computing,'' \emph{Commun. ACM}, vol.~66, no.~1, p. 74–85,
  dec 2022.

\bibitem{PATTERSON_HENNESSY_TURING}
J.~L. Hennessy and D.~A. Patterson, ``A new golden age for computer
  architecture,'' \emph{Commun. ACM}, vol.~62, no.~2, p. 48–60, jan 2019.

\bibitem{9748063}
Y.~Zhao, R.~Xie, G.~Xin, and J.~Han, ``A high-performance domain-specific
  processor with matrix extension of risc-v for module-lwe applications,''
  \emph{IEEE Trans. Circuits Syst. I: Reg. Papers}, vol.~69, no.~7, pp.
  2871--2884, 2022.

\bibitem{9763876}
R.~Paludo and L.~Sousa, ``Ntt architecture for a linux-ready risc-v
  fully-homomorphic encryption accelerator,'' \emph{IEEE Trans. Circuits Syst.
  I: Reg. Papers}, vol.~69, no.~7, pp. 2669--2682, 2022.

\bibitem{pi0}
{Raspberry Pi Foundation}, ``Rasperry pi zero,''
  \url{https://www.raspberrypi.com/products/raspberry-pi-zero}, 2020.

\bibitem{Jetson}
M.~Ditty, ``Nvidia orin system-on-chip,'' in \emph{2022 IEEE Hot Chips 34
  Symp.}, 2022, pp. 1--17.

\bibitem{unmatched}
{SiFive}, ``Hifive unmatched,''
  \url{https://www.sifive.com/boards/hifive-unmatched}, 2022.

\bibitem{ESP}
J.~Zuckerman, P.~Mantovani, D.~Giri, and L.~P. Carloni, ``Enabling
  heterogeneous, multicore soc research with risc-v and esp,'' 2022.

\bibitem{BYOC}
J.~Balkind, K.~Lim, M.~Schaffner, F.~Gao, G.~Chirkov, A.~Li, A.~Lavrov, T.~M.
  Nguyen, Y.~Fu, F.~Zaruba, K.~Gulati, L.~Benini, and D.~Wentzlaff, \emph{BYOC:
  A "Bring Your Own Core" Framework for Heterogeneous-ISA Research}.\hskip 1em
  plus 0.5em minus 0.4em\relax New York, NY, USA: ACM, 2020, p. 699–714.

\bibitem{KRAKEN}
A.~Di~Mauro, M.~Scherer, D.~Rossi, and L.~Benini, ``Kraken: A direct
  event/frame-based multi-sensor fusion soc for ultra-efficient visual
  processing in nano-uavs,'' in \emph{IEEE Hot Chips 34 Symp.}, 2022, pp.
  1--19.

\bibitem{VEGA}
D.~Rossi, F.~Conti, M.~Eggiman, A.~D. Mauro, G.~Tagliavini, S.~Mach,
  M.~Guermandi, A.~Pullini, I.~Loi, J.~Chen, E.~Flamand, and L.~Benini, ``Vega:
  A ten-core {SoC} for {IoT} endnodes with {DNN} acceleration and cognitive
  wake-up from {MRAM}-based state-retentive sleep mode,'' \emph{{IEEE} J.
  Solid-State Circuits}, vol.~57, no.~1, pp. 127--139, jan 2022.

\bibitem{HULKV_DATE}
L.~Valente, Y.~Tortorella, M.~Sinigaglia, G.~Tagliavini, A.~Capotondi,
  L.~Benini, and D.~Rossi, ``{HULK-V: a Heterogeneous Ultra-low-power Linux
  capable RISC-V SoC},'' in \emph{2023 Design, Automation \& Test in Europe
  Conference \& Exhibition (DATE)}, 2023.

\bibitem{HYPER}
{Cypress}, ``Hyperram \& octal xspi ram memory,''
  \url{https://www.cypress.com/products/hyperram-octal-xspi-ram-memory}.

\bibitem{pulpnn1}
A.~{Garofalo}, M.~{Rusci}, F.~{Conti}, D.~{Rossi}, and L.~{Benini}, ``Pulp-nn:
  A computing library for quantized neural network inference at the edge on
  risc-v based parallel ultra low power clusters,'' in \emph{26th IEEE Int.
  Conf. Electronics, Circuits and Syst.}, 2019, pp. 33--36.

\bibitem{manor1}
E.~Manor and S.~Greenberg, ``Custom hardware inference accelerator for
  tensorflow lite for microcontrollers,'' \emph{IEEE Access}, vol.~10, pp.
  73\,484--73\,493, 2022.

\bibitem{zaruba2019cost}
F.~{Zaruba} and L.~{Benini}, ``The cost of application-class processing: Energy
  and performance analysis of a linux-ready 1.7-ghz 64-bit risc-v core in 22-nm
  fdsoi technology,'' \emph{IEEE Trans. Very Large Scale Integration (VLSI)
  Syst.}, vol.~27, no.~11, pp. 2629--2640, Nov 2019.

\bibitem{opdenacker2017linux}
M.~Opdenacker, ``Linux in less than 4 mb of ram,''
  \url{https://bootlin.com/pub/conferences/2017/jdll/opdenacker-embedded-linux-in-less-than-4mb-of-ram/},
  2017.

\bibitem{AXIPULP}
A.~Kurth, W.~Ronninger, T.~Benz, M.~Cavalcante, F.~Schuiki, F.~Zaruba, and
  L.~Benini, ``An open-source platform for high-performance non-coherent
  on-chip communication,'' \emph{IEEE Trans. Comput.}, pp. 1--1, 2021.

\bibitem{arm2021amba}
A.~Limited, ``Amba axi and ace protocol specification version h.c,''
  \url{https://developer.arm.com/documentation/ihi0022/hc}, 2021.

\bibitem{regbus}
\BIBentryALTinterwordspacing
\emph{Generic Register Interface}, PULP Platform Contributors. [Online].
  Available: \url{https://github.com/pulp-platform/register_interface}
\BIBentrySTDinterwordspacing

\bibitem{benz2023highperformance}
T.~Benz, M.~Rogenmoser, P.~Scheffler, S.~Riedel, A.~Ottaviano, A.~Kurth,
  T.~Hoefler, and L.~Benini, ``A high-performance, energy-efficient modular dma
  engine architecture,'' 2023.

\bibitem{RPC-spec}
{Etron Technology}, ``256{Mb} high bandwidth rpc dram,''
  \url{https://etronamerica.com/wp-content/uploads/2019/05/EM6GA16LGDABMACAEA-RPC-DRAM_Rev.-1.0.pdf},
  2019.

\bibitem{LITEDRAM}
{Antmicro}, ``Open source ddr controller framework for mitigating rowhammer,''
  \url{https://antmicro.com/blog/2021/08/open-source-ddr-test-framework-for-rowhammer/},
  2021.

\bibitem{TUK}
C.~Sudarshan, J.~Lappas, C.~Weis, D.~M. Mathew, M.~Jung, and N.~Wehn, ``A lean,
  low power, low latency dram memory controller for transprecision computing,''
  in \emph{Embedded Comput. Syst.: Architectures, Model., Simul.}\hskip 1em
  plus 0.5em minus 0.4em\relax Springer Int. Publishing, 2019, pp. 429--441.

\end{thebibliography}
\end{document}